\documentclass[aps,prb,showpacs]{revtex4}
\input epsf.tex
\begin{document}

\title{The Correlated Kondo-lattice Model}

\author{{\sc J. Kienert}, {\sc C. Santos}, and {\sc W. Nolting}} 

\address{Institut f\"ur Physik, Humboldt-Universit\"at zu Berlin,
Newtonstra{\ss}e 15, 12489 Berlin, Germany}



\date{\today}

\begin{abstract}
  We investigate the ferromagnetic Kondo-lattice model (FKLM) with a 
  correlated conduction band. A moment conserving approach is proposed
  to determine the electronic self-energy. Mapping the interaction onto
  an effective Heisenberg model we calculate the ordering of the localized
  spin system self-consistently. Quasiparticle densities of states
  (QDOS) and the Curie temperature $T_{\rm C}$ are calculated. The
  band interaction leads to an upper Hubbard peak and modifies
  the magnetic stability of the FKLM. 
\end{abstract}

\pacs{71.27.+a, 75.30.Mb, 75.30.Vn}

\maketitle

\section{Introduction}

There has been renewed interested in the ferromagnetic Kondo-lattice
model \cite{ZEN51, AH55} since the discovery of the 
colossal magnetoresistance (CMR) \cite{RAM97,FUR98,EG01}. This model
consists of uncorrelated (s-)band electrons that interact
intra-atomically with localized quantum spins (Hund's rule
coupling). Apart from model extensions such 
as electron-phonon interaction \cite{EG01}, the role of electronic
correlations among the conduction electrons has been emphasized
\cite{HV00}. These correlations are often incorporated as a Hubbard-like
interaction, i.\,e. the conduction electrons interact locally via a
repulsive Coulomb matrix element $U$ \cite{HUB63}. The Hamiltonian of
the {\em correlated} FKLM reads (for simplicity we assume one
($s$-)orbital):  
\begin{equation}
\label{eq:H}
  H=\sum_{ij\sigma}T_{ij}c_{i\sigma}^{\dag}c_{j\sigma}-J\sum_{i}\mbox{\boldmath${\sigma}$}_{i}\cdot {\bf S}_{i}+\frac{1}{2}U\sum_{i\sigma}n_{i\sigma}n_{i-\sigma}\;.  
\end{equation}
As usual, $c_{i\sigma}^{\dag}~(c_{i\sigma})$ creates (annihilates)
an electron of spin $\sigma~(=\uparrow,\downarrow)$ at site $i$, $T_{ij}$
are the hopping integrals, and \boldmath${\sigma}$\unboldmath$_{i}$ and 
${\bf S}_{i}$ denote the conduction electron spin and the localized spin,
respectively, coupled by an intra-atomic exchange constant $J$. The
notion ferromagnetic is due to a positive $J$. The (F)KLM is
also known as s-d or s-f model; if the (positive) Hund coupling greatly
exceeds the kinetic energy, the double exchange model is obtained
\cite{ZEN51, AH55}.    

With estimated values of the bandwidth $W\simeq 1-2$
eV, $J\simeq 1$ eV, and $U\simeq 8$ eV for LaMnO$_{3}$, as given in \cite{SPV96}, the
Hubbard interaction has certainly to be taken into account.  
A theory of the (ferromagnetic) {\it correlated} Kondo-lattice model
should therefore treat both the Hund and the Hubbard interaction as
strong couplings. Furthermore, we stress the importance of the quantum
character of the localized spins, which is usually neglected
\cite{FUR98,HV00}. On the one hand,
assuming classical (localized) spins allows for the application of
Dynamical Mean Field Theory, a state-of-the-art theory for strongly
correlated systems; on the other hand, the influence of electron-magnon
interaction ("spin-flips"), which is suppressed when using classical
spins, on the electronic spectrum is rather considerable
\cite{SN02}.           

We present a theory that accounts for the strong coupling
aspect as well as quantum mechanical spins. First, a moment conserving
decoupling procedure (MCDA), which has already been applied to the
uncorrelated Kondo-lattice model both for bulk and film geometries
\cite{SN02,ScN01}, yields the electronic self-energy. 
Secondly, the Hund coupling between localized spins and
conduction electrons is mapped onto an effective Heisenberg
operator by integrating out the electronic degrees of freedom
\cite{SN02}. 
   
As an extension to the theory for the uncorrelated Kondo-lattice model,
our approach incorporates the electron-electron 
interaction according to the decoupling scheme proposed by Hubbard
himself ("Hubbard-I") \cite{HUB63}. The modified RKKY interaction is
formulated by means of an effective medium method to take care of the
additional correlations.

\section{Theory}
For technical details the reader is referred to \cite{SN02}. The
equation of motion (EOM) for the Green's function 
$G_{ij\sigma}(E)=\langle\langle c_{i\sigma};c_{j\sigma}^{\dag}
\rangle\rangle_{E}$ yields 
\begin{equation}
  \sum_{l}\left(E\delta_{il}-T_{il}\right)G_{lj\sigma}(E)=\hbar\delta_{ij}+U\Gamma_{iii,j\sigma}(E)-\frac{1}{2}J\left\{z_{\sigma}I_{ii,j\sigma}(E)+F_{ii,j\sigma}(E)\right\},
\end{equation}
with $z_{\uparrow}=1,~z_{\downarrow}=-1$, and the higher Green's functions
\begin{equation}
   {\Gamma}_{ikl,j\sigma}(E) = \langle\langle{
     c_{i-\sigma}^{\dag}c_{k-\sigma}c_{l\sigma};c^{\dag}_{j\sigma}}\rangle\rangle_{E}\;,
\end{equation}
\begin{equation}
  I_{ik,j\sigma}(E)=\langle\langle{S_{i}^{z}c_{k\sigma};c^{\dag}_{j\sigma}}\rangle\rangle_{E}~,
~~~~~  F_{ik,j\sigma}(E)=\langle\langle{S_{i}^{-\sigma}c_{k-\sigma};c^{\dag}_{j\sigma}}\rangle\rangle_{E}\;.
\end{equation}
One proceeds with writing down again the equations of motion and
decouples the resulting still higher Green's
functions. Fortunately, within the frame of the Hubbard-I decoupling, the
Hubbard-Green's function can be expressed as a functional of already
known quantities: 
\begin{equation}
\Gamma_{iii,j\sigma}(E)={\cal
  G}\left[G_{ij\sigma}(E),I_{ii,j\sigma}(E),F_{ii,j\sigma}(E)\right]\;.   
\end{equation}
The higher Green's functions in the equations of motion of
$I_{ik,j\sigma}(E)$ and $F_{ik,j\sigma}(E)$, too, are projected onto
$G,\;I,\;{\rm and\;}F$ and in addition, where electron density
correlations show up explicitly, onto $\Gamma$. The
coefficients are fixed via sum rules. From this closed system of
equations, one obtains an electronic self-energy of the following
structure: 
\begin{equation}
\label{eq:selfenergy}
    \Sigma_{ij\sigma}(E)={\cal S}\left[\Sigma_{ij\sigma}(E),\langle
      n_{i\sigma} 
    \rangle,\langle S_{i}^{z} \rangle,\langle (S_{i}^{z})^{2}
      \rangle,\langle (S_{i}^{z})^{3} \rangle,\langle
      S_{i}^{\pm}S_{i}^{\mp}\rangle,...\right]\;. 
\end{equation}

The mapping of the Hund's rule interaction
onto an effective Heisenberg-like spin-spin operator
$H_{SS}$ is achieved by averaging it in the subspace of the
$s$-electrons. 
In our case this is done with an
effective Hamiltonian \cite{MN99} that incorporates the Hubbard
interaction as a renormalized kinetic term. The result is an anisotropic
Heisenberg Hamiltonian:
\begin{equation}
\label{eq:Hss}
H_{SS}=-\sum_{i,j} \left\{{\hat J}^{(1)}_{ij}\left(
      S_{i}^{+}S_{j}^{-}+S_{i}^{-}S_{j}^{+}\right)+{\hat
      J}^{(2)}_{ij}S_{i}^{z}S_{j}^{z}\right\}+B_{\rm eff}\sum_{i}S_{i}^{z}\;.
\end{equation}   
The effective exchange integrals 
\begin{equation}
  {\hat J}^{(1)}({\bf q})=-\frac{1}{8}J^2{\cal D}_{\bf
    q}^{\uparrow\downarrow}\;,
 ~~~~ {\hat J}^{(2)}({\bf q})=-\frac{1}{8}J^2\sum_{\sigma}{\cal D}_{\bf q}^{\sigma\sigma} 
\end{equation}
become temperature-dependent via 
\begin{equation}
  {\cal D}_{\bf q}^{\sigma\sigma^{\prime}}=-\frac{1}{\pi N}\Im
  \int_{-\infty}^{+\infty}{\rm d}E f_{-}(E)\sum_{\bf k} \left(G_{{\bf k}\sigma^{\prime}}^{(U)}(E)G_{{\bf k}+{\bf
      q},\sigma}(E)+G_{{\bf k}+{\bf
      q},\sigma}^{(U)}(E)G_{{\bf k}\sigma^{\prime}}(E)\right).
\end{equation}
$\Im$ denotes the imaginary part, $f_{-}(E)$ is the Fermi function, and 
\begin{equation}
  G_{{\bf k}\sigma}^{(U)}(E)=\frac{\hbar}{E-\epsilon({\bf
      k})-\Sigma^{(U)}_{{\bf k}\sigma}(E)}\;, B_{\rm
    eff}=\frac{J}{2\pi N} \Im
    \int_{-\infty}^{+\infty}{\rm d}E f_{-}(E)\sum_{\bf
      k \sigma}z_{\sigma}G_{{\bf k}\sigma}^{(U)}(E),
\end{equation}
where $\Sigma^{(U)}_{{\bf k}\sigma}(E)$ is an effective medium
"Hubbard-self-energy part". For $U=0$ and $J\ll t$, one recovers the
conventional RKKY-interaction \cite{SN02}. The spin expectation values in 
Eq. (\ref{eq:selfenergy}) are obtained by applying a
Tyablikov-decoupling to the EOM of a Green's function according to Callen
(for arbitrary $S$) built with $H_{SS}$
\cite{SN02,Cal63}. The same method yields a rather simple formula
to calculate the Curie temperature of the system:
\begin{equation}
  k_{\rm B}T_{\rm C}\simeq\frac{1}{3}S(S+1)\frac{1}{\frac{1}{N}\sum\limits_{\bf
      q}\left(\frac{B_{\rm eff}}{\langle S^{z}\rangle}+2({\hat
      J}^{(2)}({\bf 0})-2{\hat J}^{(1)}({\bf q}))\right)^{-1}_{T \simeq T_{c}}}\;.
\end{equation}

\section{Results and Discussion}
\begin{figure}[hbtp]
  \vspace{-9pt}

  \centerline{\hbox{ \hspace{0.0in} 
    \epsfxsize=2.5in
    \epsffile{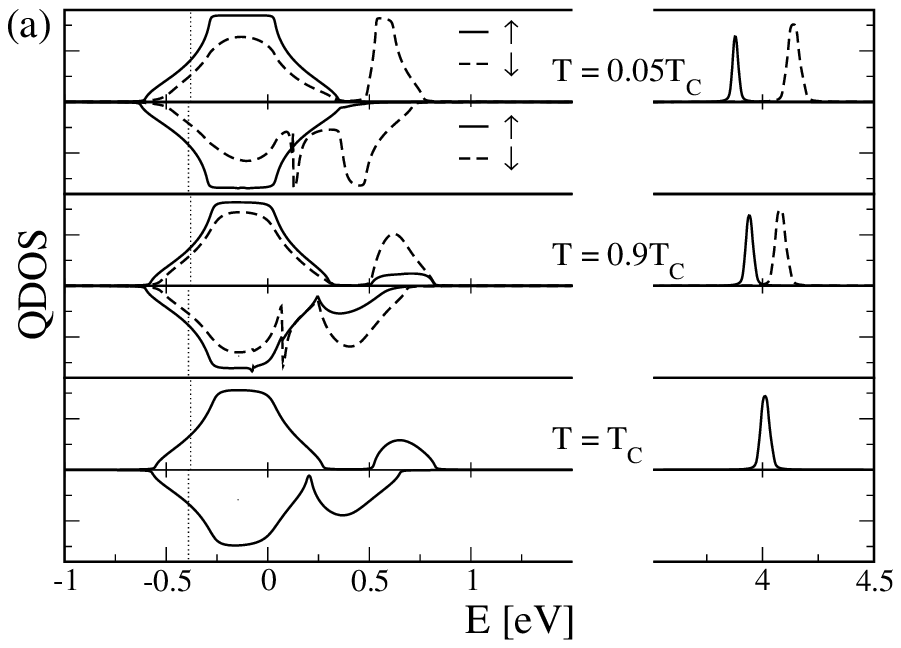}
    \hspace{0.1in}
    \epsfxsize=2.5in
    \epsffile{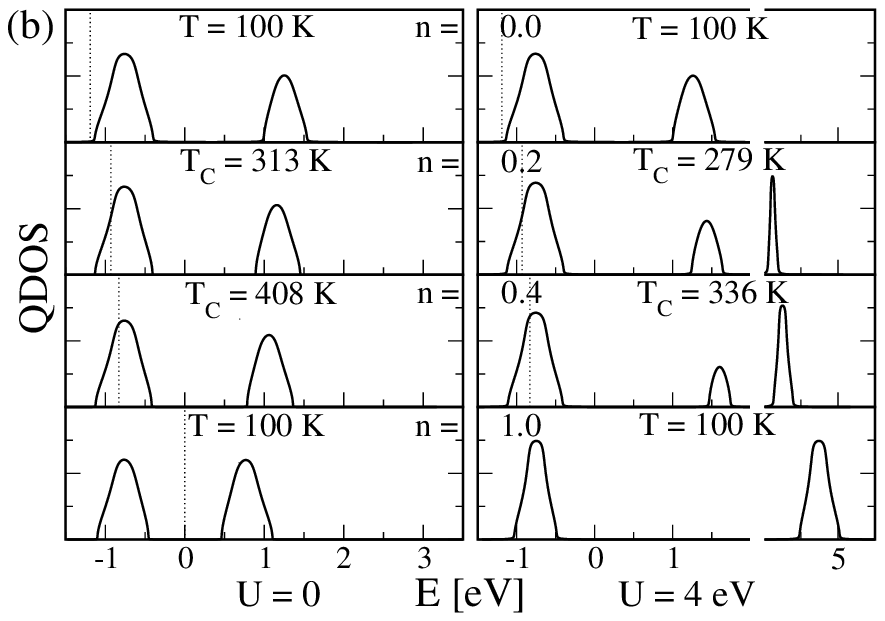}
    }
  }
  \caption{Quasiparticle densities of states. (a) Upper part: $U$\,=\,4\,eV,
  $T_{\rm C}=144$\,K, lower part: $U=0$, $T_{\rm C}=88$\,K. Other parameters:
  $n=0.15$, $J=0.5$\,eV, $S=\frac{1}{2}$. (b) Parameters: $J=1$\,eV,
  $S=\frac{3}{2}$. For $n=0$ and $n=1$ the temperature is
  arbitrary as the system is not self-consistently ferromagnetic.}  
  \label{fig1}
\end{figure}

We investigated the numerical results of our equations on a simple cubic
lattice with bandwidth $W=1$\,eV, restricting ourselves to ferromagnetism.

In Fig. \ref{fig1}(a) the QDOS is shown at key temperatures
in the correlated ($U=4\;$eV) and uncorrelated ($U=0$) case. Note the
considerable amount of 
$\downarrow$-spectral weight even in the ferromagnetically
quasi-saturated case (which will not disappear for large $J$, but rather
go into saturation on a non-negligible level). The main effect of a finite
Hubbard interaction is a removal of spectral weight from the low-energy
region, thus creating a gap between the lower subband and the
polaron-like second subband. The center of gravity of the latter is
shifted to higher energies with increasing $U$. We consider the
Stoner-like splitting of the upper Hubbard-band at $E\sim U+\frac{1}{2}JS$
in Fig. \ref{fig1}(a) as an artefact of our approximate theory.

Figure \ref{fig1}(b) shows QDOS for different values of the band
occupation in the paramagnetic regime. For finite $U$ and intermediate
band occupation, a correlation-induced three-band structure is clearly
visible. In the uncorrelated case one observes a shift of the upper band
with increasing $n$, changing its character from a polaron-like band
(low $n$) to a double-occupation band. Note also the smaller bandwidth
in the low energy region for $U=4$\,eV due to the reduced effective
hopping of the electrons.      

\begin{figure}[hbtp]
  \centerline{\hbox{ \hspace{0.0in} 
    \epsfxsize=2.5in
    \epsffile{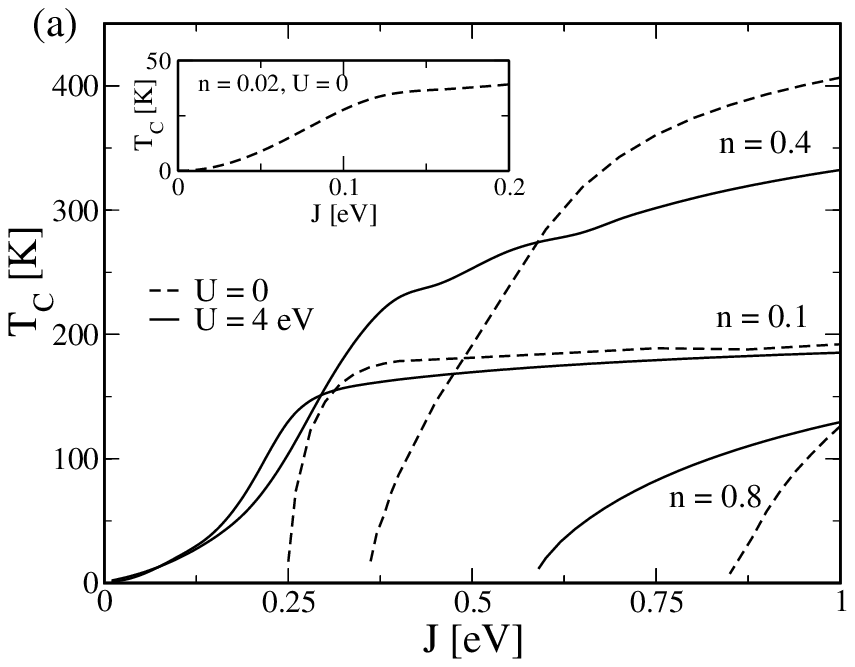}
    \hspace{0.1in}
    \epsfxsize=2.5in
    \epsffile{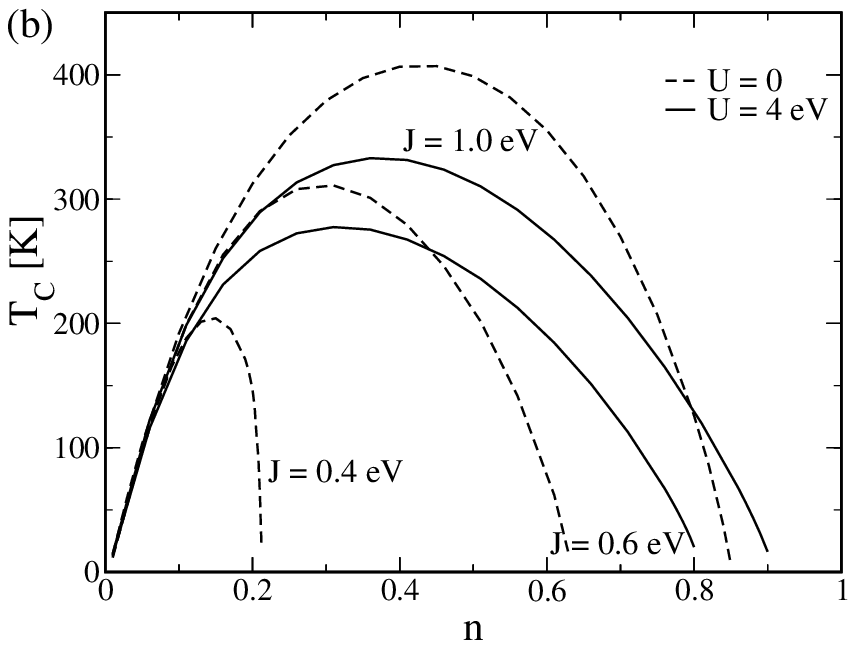}
    }
  }
  \caption{Curie temperatures of the (correlated)
    Kondo-lattice model ($S=\frac{3}{2}$) as a function of (a) the
    Hund coupling $J$ and (b) the band occupation $n$.}
  \label{fig2}
\end{figure}

Curie temperatures are displayed in Fig. \ref{fig2}. The inset in
Fig. \ref{fig2}(a) shows the conventional RKKY-like behaviour $T_{\rm
  C}\sim J^2$ for small $J$ and $n$. A characteristic feature of the modified
RKKY theory is, for $n>n_{\rm c}$, a critical value of the Hund
coupling $J_{\rm c}$ below which there is no
ferromagnetism. It is interesting to note that the critical density $n_{\rm c}$ coincides
with the magnetic phase boundary when using conventional RKKY 
theory. By switching on electronic correlations, we see that the
critical interaction $J_{\rm
  c}$ is shifted to lower values, particularly restoring the system's
ability to exhibit RKKY-ferromagnetism.
Furthermore, $T_{\rm C}(J)$ runs into saturation  
\cite{SN02}. For $J\gg t$ (in the double exchange regime),
there is suppression of ferromagnetic stability due to the Hubbard
interaction: as in this regime $T_{\rm C}\sim t$, and a finite
$U$ reduces the kinetic energy of the conduction band electrons, $T_{\rm
  C}$ decreases.  

As can be seen in the $T_{\rm C}-n$ phase diagram
of Fig.\,\ref{fig2}(b), the ferromagnetic regime is extended to
higher band occupations. However, we did not get any ferromagnetism
neither for the uncorrelated nor for the correlated half-filled
band. This is consistent with other results \cite{WAN98} and with
experiment \cite{RAM97}, where $n \simeq 1$ 
corresponds to low doping. It goes without saying that our
model study does not allow for a detailed comparison with real
CMR-materials. Electron-phonon coupling and orbital degeneracy would
have to be included, as well as antiferromagnetic calculations to be
done. However, the Curie temperature has the right order of magnitude,
and runs through a maximum when changing the doping, as observed in
\cite{RAM97}.  

In conclusion, we have presented a fully self-consistent theory of the
correlated (ferromagnetic) Kondo-lattice model with quantum mechanical
spins. The electronic spectrum exhibits a correlation-induced multi-band
structure due to the Hubbard interaction, which also modifies the
behaviour of the critical temperature. The pronounced suppression of
$T_{\rm C}$ for intermediate band occupation (doping) and large Hund
coupling indicates that Coulomb correlations should not be neglected
when modelling CMR substances.     

\end{document}